# Optothermal Molecule Trap

*Stefan Duhr and Dieter Braun[+]*

[+]*Biophysics, Ludwig Maximilians Universität München,
Amalienstr. 54, 80799 München, Germany
e-mail: dieter.braun@physik.lmu.de, mail@dieterb.de*

**Thermophoresis moves molecules along temperature gradients, typically from hot to cold. We superpose fluid flow with thermophoretic molecule flow under well defined microfluidic conditions, imaged by fluorescence microscopy. DNA is trapped and accumulated 16-fold in regions where both flows move in opposite directions. Strong 800-fold accumulation is expected, however with slow trapping kinetics. The experiment is equally described by a three-dimensional and one-dimensional analytical model. As an application, we show how a radially converging temperature field confines short DNA into a 10 μm small spot.**

*Introduction.* Contact free methods to manipulate single particles are rare. Common tools are electrophoresis and optical-tweezers. However, in optical trapping, forces scale with particle volume, limiting the method to particles larger than about 500 nm. Less well known as a tool for optical molecule manipulation is thermophoresis[1,2], where molecules are moved by a thermal gradient. With laser heating, it allows all-optical microscale access to small molecules, in contrast to electrophoresis. The achieved active accumulation of biomolecules helps to overcome diffusion limitations in both surface and bulk biochemical reactions. A well known example is the diffusion limited reaction kinetics on DNA microarrays.

Recent advances in measuring thermophoresis using holographic scattering[3-6], optical lensing[7-9], beam deflection[10-12] and optothermal microfluidics[13,14] lead to a better understanding of thermophoresis[15-19]. Most experiments to measure thermophoresis[3,9,10-14] use conditions where fluid flow can be neglected. However, in elongated columns thermophoretic accumulation can be amplified by thermal convection[20-23], and flat geometries with toroidal convection flow yield point accumulations[13,28]. In thermal field flow fractionation (ThFFF), the fluid flows perpendicular to the temperature gradient and separates molecules by their respective thermophoretic response[24-26]. For the first time, we orient the fluid flow along the thermal gradient, made possible with optical heating by an infrared laser focus. The combination of thermophoresis and fluid flow results in strong trapping of small biomolecules (Fig. 1). As we will show, the experiments are equally described by three-dimensional (3D) or one-dimensional (1D) models.

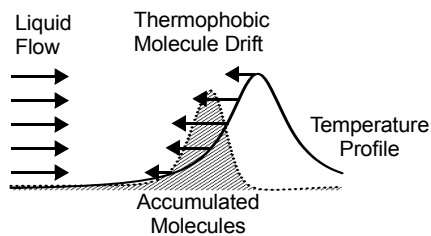

Fig. 1. *Principle of Thermophoretic Flow Trap*. A warm spot repels molecules by thermophoresis. Counteracting fluid flow leads to accumulation of molecules upstream of the warm spot.

*Flow trap.* Molecules drift away from the heat by thermophoresis, counteracted by a passive liquid flow. As a result, molecules are trapped upstream of the warm spot. Generally accepted is a phenomenological foundation of the thermophoretic drift based on the Onsager theory of linear non equilibrium thermodynamics. For low molecule concentrations the thermophoretic drift velocity v is proportional to temperature gradient $\nabla T$:

$$\vec{v_T} = -D_T \nabla T \tag{1}$$

The coefficient $D_T$ is termed thermodiffusion coefficient with units $m^2/(sK)$ analog to the electrophoretic mobility $\mu$ with units $m^2/(sV)$. Thermophoresis extends Fick's law to

$$\vec{j} = -D\nabla c - (1-c)D_T c \nabla T \tag{2}$$

with molecule concentration c and diffusion coefficient D. In steady state, with fluid at rest, diffusional and thermophoretic flow balance. Fluid flow directly opposes a thermal gradient in a microfluidic channel with 10μm x 10μm cross section (Fig. 2a). The channel is surrounded by PDMS (Polydimethylsiloxane) silicone using a manufacture protocol described previously[14]. We oppose the flow of DNA containing water with a locally enhanced temperature gradient, created with a focussed infrared laser (Furukawa FOL1405-RTV-317, 1480 nm). Fluid flow is controlled gravitationally by two open 5 μl syringes of defined height, connected to the microfluidics with silicone tubings (Carl Roth Laborbedarf).

The channel is imaged with a 40x oil objective on an AxioTech Vario fluorescence microscope (Zeiss), illuminated with a high power LED (Luxeon) and recorded with the CCD Camera SensiCam QE (PCO). Details of bleaching correction and temperature extraction were described previously[14]. Temperature increase is measured by the temperature dependent fluorescence signal of the dye BCECF, diluted to 50 μM in 10 mM TRIS buffer. From the total BCECF temperature dependence of -2.4 %/K, only -0.9 %/K stems from pH drift of the used TRIS buffer. The remaining -1.5 %/K are the result of thermophoresis of the dye itself[14]. Temperature was increased by 16 K in a Lorentzian-shaped focus $T = T_0 + \Delta T/(1 + (x/b)^2)$ of width b = 12.2 μm (Fig. 2b). DNA concentration was measured by fluorescence using an intercalating dye at low concentration. Highly monodisperse and pro-tein-free DNA of 10000 bp (Fermentas) was fluorescently labeled by using 1x SYBR-I (Molecular Probes, Oregon) which shows 1000x fluorescence increase when bound to DNA. DNA was diluted to 10 nM in 1 mM TRIS buffer

DNA is flowing from the left with an effective velocity of 0.55 µm/s against the warm spot. Within 15 minutes, DNA accumulates 16-fold in the front, i.e. left of the warm spot. Images of DNA concentration recorded by fluorescence at different times are shown in Fig. 2d. First, DNA is depleted at the warm spot (right side), followed by accumulation in front of the depletion (left side) as the water drift from the left brings in DNA. The depletion on the right is slowly fading in the progress of accumulation, filled by diffusion from the accumulated DNA. The effective fluid velocity was measured by the speed with which the accumulated DNA drifted to the right after the laser was switched off.

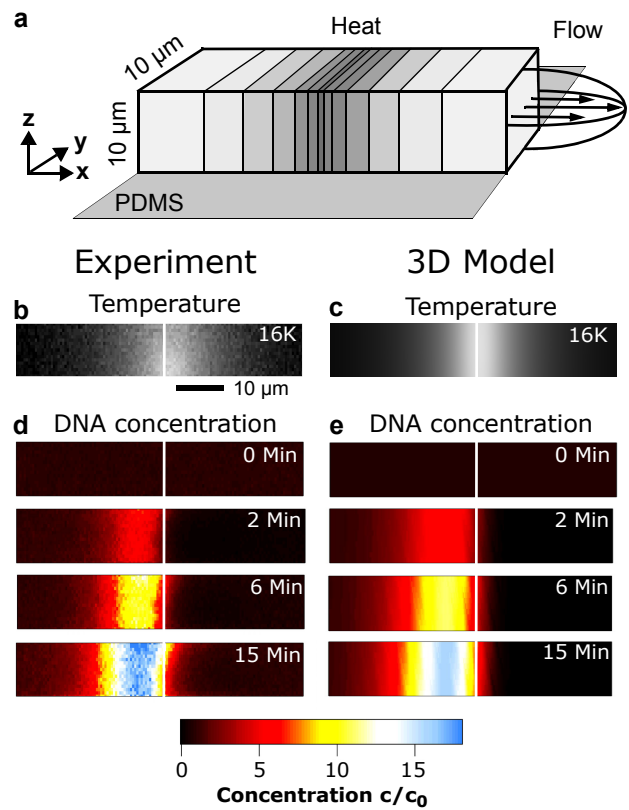

Fig. 2. ***Thermophoretic flow trap***. (a) A warm spot (gray scale; $\Delta T_{Max}$ = 16K ) repels molecules by thermophoresis. Counteracting parabolic fluid flow (black arrows; $v_{Max}$ = 0.55 µm/s) leads to accumulation of molecules upstream of the warm spot. (b) Cross chamber average of fluorescently measured temperature. (c) 3D temperature simulation of a heated water film between PDMS sheets. Low heat conductivity of PDMS leads to homogenous temperature across the channel (d) The experiment shows that DNA is strongly accumulated 16 fold within 15 min. (e) 3D simulation of the microfluidic chamber using a parabolic flow profile. Kinetics and magnitude of accumulation fit the experiment.

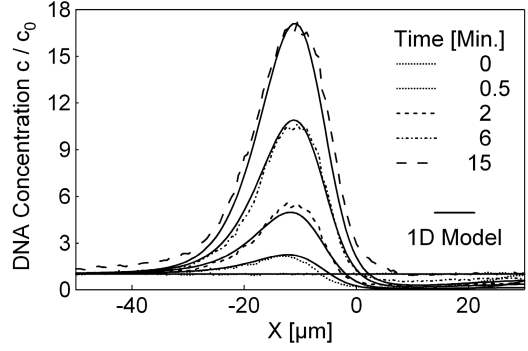

Fig. 3. *1D simulation describes experimental results*. (a) The experimental cross chamber average (dotted lines) match with theoretical results of a 1D time dependent simulation.

*Three-dimensional model.* We first model the accumulation in three dimensions with a parabolic flow $\vec{v}(y,z)$ with a peak velocity of 0.55 µm/s. The extended Ficks' law is given by:

$$\vec{j} = \vec{v}(y,z)c - D\nabla c - D_T c \nabla T \qquad (3)$$

We used the experimentally obtained Lorentzian temperature profile with a constant temperature in y- and z-direction due to low thermal conductivity of PDMS (Fig. 2b,c). Diffusion coefficient $D = 1.7$ µm$^2$/s and thermophoretic mobility $D_T = 1.1$ µm$^2$/(sK) were taken from measurements on 10.000 base pair DNA in 1 mM TRIS buffer[15]. Convection has proven insignificant in these thin fluid films[14] with convection velocities on the order of nm/s. We plot the simulated cross section DNA concentrations over time next to the corresponding experimental results (Fig. 2d,e). As expected, due to the fast diffusion relaxation time across the channel on the order of $\tau = \Delta x^2/D = 0.2$ Min., the spacial inhomogenous velocity profile $\vec{v}(y,z)$ is smeared out in the accumulation and does not lead to a spacial inhomogeneity.

*One-dimensional model.* The model can be reduced to one dimension due to fast diffusion across the channel. We will show that it describes the experiments equally well. We assume a plug flow instead of an parabolic flow profile $\vec{v}(y,z) = v_{Max}\vec{e}_x$. Fig. 3 shows the DNA concentrations at the center of the channel at various times. The overlaid solid lines from the one dimensional simulation fit the data very well. We see that thermophoresis opposed by a fluid flow can be described by a straightforward superpositions of flows by Eq.(3). It leads to a fast and strong accumulation of DNA molecules. Accumulations at longer times in steady state can be described by an analytical model. Without fluid flow, thermophoresis and diffusion balance in steady state. For constant D and $D_T$ at low DNA concentration c compared to water molecules Eq.(2) can be integrated to

$$c(x) = c_0 e^{-U(x)} \qquad U(x) = \frac{D_T[T(x) - T_0]}{D} \tag{4}$$

with boundary values for concentration $c_0$ and temperature $T_0$. Formally, the term $U(x)$ can be interpreted as thermodynamic potential. The local concentration c becomes a direct function of local temperature T, irrespective of the steepness of the temperature gradient $\nabla T$. With fluid flow, the continuity equation $\partial c / \partial t + \nabla \cdot j = 0$ has to be taken into account. At steady state $\partial c / \partial t = 0$, the molecule flow j is constant given by $j_0 = v c_0$ with drift velocity v. An analytical solution of the steady state concentration profile c(x) can be integrated:

$$c(x) = \frac{v}{D} e^{-U(x)} \int_x^\infty e^{U(x')} dx'$$

$$U(x) = \frac{D_T[T(x) - T_0] - vx}{D} \tag{5}$$

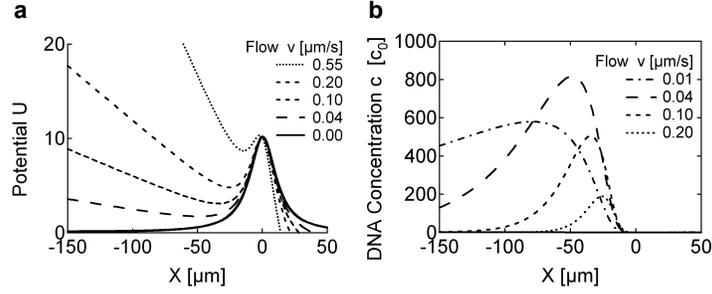

Fig. 4. *Steady state accumulation.* (a) In the potential image, introduction of flow corresponds to a tilting of thermophoretic depletion potential U, leading to a potential trough upstream of the warm spot. (b) The minimum of effective potential coincides with the maximum concentration in steady state. Accumulation peaks at 820x for v=0.04 μm/s.

The term $U(x)$ can be interpreted as an effective potential. The normalization integral in equation (5) does not affect the shape of the accumulation profile near its peak. The potential allows a intuitive description of the accu-mulation: as the drift velocity increases, the initially depleting potential U for v = 0 builds up a potential trough upstream (left) of the warm spot (Fig. 4a). The trough dimensions become smaller as the drift velocity v increases and is large in width and depth for small flow velocities. Based on the potential image, we understand the strong attraction of molecules in steady state calculated for low values of v (Fig. 4b). The accumulation diminishes as drift v increases and the tilting of the potential by the term $-vx/D$ leads to a shallower and less wide potential trough. For the steady state solution in a one-dimensional model, accumulation is peaking for slower drift velocities. At v=0.04 µm/s, accumulation is maximal with 820-fold DNA concentration in front of the heat spot (Fig. 4b). While smaller drift v makes the potential deeper and accumulation stronger (Fig. 4a,b), considerably more time is needed to transport the DNA to the trap and reach a steady state. For example, the time to reach above 820x accumulation steady state is 1000 h or 41 days. The peak concentration is located with good approximation at the minimum of the effective potential U(x), i.e. the maximal concentration is located at the point where thermophoretic drift $v_T = -D_T \nabla T$ and flow drift v exactly oppose each other:

$$\nabla U = 0 \quad \Rightarrow \quad v = D_T \nabla T \tag{6}$$

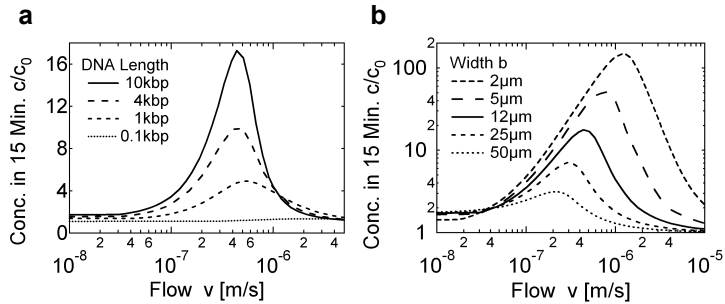

Fig. 5. *Accumulation in finite time*. (a) Maximal accumulation in finite time requires faster drift v than in steady state. The optimal velocity of 0.55 μm/s is only slightly shifted for shorter DNA. (b) Tighter focussing makes the potential steeper and leads to faster equilibration and considerably stronger accumulation within finite time.

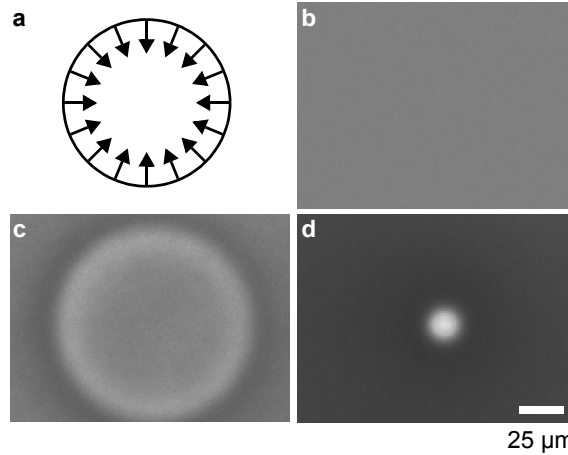

Fig. 6. **Radial DNA accumulation.** (a) A radial temperature circle is written into the chamber. It is closed radially over time and DNA becomes accumulated in the center. (b-d) Sequence of concentration of 1000 bp DNA at times 0 s, 20 s and 80 s. Finally DNA is 6-fold accumulated in a spot 10 µm in radius.

*Flow and gradient dependence.* The point of maximum accumulation is located closer to the heated center for higher velocities v (Fig. 4a,b), since the temperature gradient is stronger there. Fig. 5a shows the velocity dependence of DNA accumulation within a short time of 15 min. To reach maximal accumulation of 10 kbp DNA within this time, v=0.55 µm/s is optimal. Both faster and slower drift decreases the accumulation in the chosen finite time. The velocity for maximal accumulation efficiency barely depends on DNA length, allowing to infer DNA size from the achieved accumulation at a constant drift v. As Fig. 5b shows, size selectivity and the magnitude of accu-mulation within the limited time span of 15 min, is exceptionally sensitive to focus width. Decreasing the focal width from 12 µm to 2 µm increases the accumulation of 10 kbp DNA within 15 min by an order of magnitude.

*Radial flow trap in a sheet of liquid.* As temperature is created by optical means, we are not limited to a fixed focus in a microfluidic flow channel. For example, DNA in a two-dimensional 20 µm high water sheet can be accumulated by optically imposed circular heat rings which move concentrically towards a spot (Fig. 6). With a radial drift of v = 2.3 µm/s and a average temperature of $\Delta T_{Max}$ = 20 K, short 1000 base pair DNA was accumulated 6-fold in a spot of 10 µm radius within 80 seconds. This result illustrates how optically triggered thermophoretic traps can accumulate molecules in a variety of geometries.

*Conclusion.* Opposing thermophoretic flow against a fluid flow creates a molecule trap. Fluorescence measurements in a microfluidic channel agrees well with the one dimensional superposition of both flows. Within 15 minutes, 10.000 base pair DNA was accumulated 16-fold by a 16 K warm microscopic spot. We showed that 1000 base pair DNA can be 6-fold accumulated to a spot by radially moving a circular temperature ring.

We thank Franz Weinert for assistance, Peter Fromherz for suggestions and Hermann Gaub for hosting the Emmy-Noether Nachwuchsgruppe which was funded by the Deutsche Forschungsgemeinschaft.

Fig. 1.  *Principle of Thermophoretic Flow Trap*. A warm spot repels molecules by thermophoresis. Counteracting fluid flow leads to accumulation of molecules upstream of the warm spot.

Fig. 2. **Thermophoretic flow trap**. (a) A warm spot (gray scale; $\Delta T_{Max} = 16K$) repels molecules by thermophoresis. Counteracting parabolic fluid flow (black arrows; $v_{Max} = 0.55$ µm/s) leads to accumulation of molecules upstream of the warm spot. (b) Cross chamber average of fluorescently measured temperature. (c) 3D temperature simulation of a heated water film between PDMS sheets. Low heat conductivity of PDMS leads to homogenous temperature across the channel (d) The experiment shows that DNA is strongly accumulated 16 fold within 15 min. (e) 3D simulation of the microfluidic chamber using a parabolic flow profile. Kinetics and magnitude of accumulation fit the experiment.

Fig. 3.  **1D simulation describes experimental results**. (a) The experimental cross chamber average (dotted lines) match with theoretical results of a 1D time dependent simulation.

Fig. 4. **Steady state accumulation.** (a) In the potential image, introduction of flow corresponds to a tilting of thermophoretic depletion potential U, leading to a potential trough upstream of the warm spot. (b) The minimum of effective potential coincides with the maximum concentration in steady state. Accumulation peaks at 820x for v=0.04 µm/s.

Fig. 5. **Accumulation in finite time**. (a) Maximal accumulation in finite time requires faster drift v than in steady state. The optimal velocity of 0.55 µm/s is only slightly shifted for shorter DNA. (b) Tighter focussing makes the potential steeper and leads to faster equilibration and considerably stronger accumulation within finite time.

Fig. 6. **Radial DNA accumulation.** (a) A radial temperature circle is written into the chamber. It is closed radially over time and DNA becomes accumulated in the center. (b-d) Sequence of concentration of 1000 bp DNA at times 0 s, 20 s and 80 s. Finally DNA is 6-fold accumulated in a spot 10 µm in radius.